\documentclass[aip,jap,reprint,twocolumn]{revtex4-1}
\usepackage[utf8]{inputenc}
\usepackage{amsmath,amssymb,amsfonts}
\usepackage{graphicx,epsfig}
\usepackage{array}
\usepackage{dcolumn}

\usepackage[squaren]{SIunits}

\begin{document}

\title{Half-metallic magnetism and the search for better spin valves}

\author{Karin Everschor-Sitte}
\author{Matthias Sitte}
\author{Allan H.~MacDonald}
\affiliation{The University of Texas at Austin, Department of Physics, 2515 Speedway, Austin, TX 78712, USA}

\begin{abstract}
We use a previously proposed theory for the temperature dependence of
tunneling magnetoresistance to shed light on ongoing efforts to
optimize spin valves.  First we show that a mechanism in which spin
valve performance at finite temperatures is limited by uncorrelated
thermal fluctuations of magnetization orientations on opposite sides of
a tunnel junction is in good agreement with recent studies of the
temperature-dependent magnetoresistance of high quality tunnel
junctions with MgO barriers.  Using this insight, we propose a simple
formula which captures the advantages for spin-valve optimization of
using materials with a high spin polarization of Fermi-level tunneling
electrons, and of using materials with high ferromagnetic transition
temperatures.  We conclude that half-metallic ferromagnets can yield
better spin-value performance than current elemental transition metal
ferromagnet/MgO systems only if their ferromagnetic transition
temperatures exceed $\sim\unit{950}{\kelvin}$.
\end{abstract}

\date{\today}

\maketitle

\section{Introduction}

\begin{figure}[tb]
\includegraphics[width=0.4\textwidth]{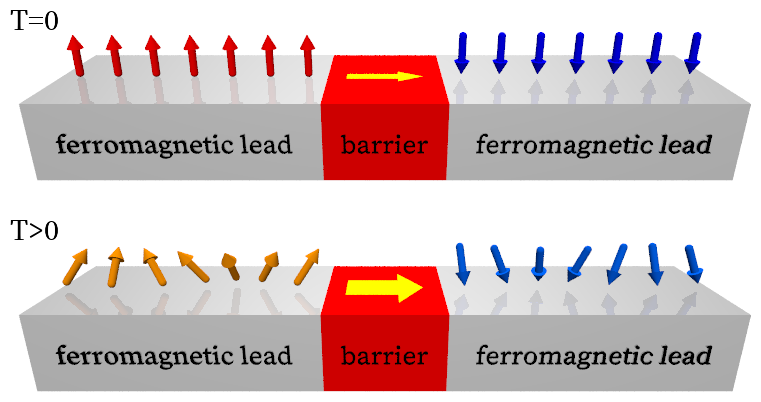}
\caption{%
(Color online) Schematic illustration of a magnetic tunnel junction
(MTJ).  In the high resistance state the magnetizations of the left and
right magnetic films at $T=0$ are exactly opposite (top panel) and the
exchange potential experienced by transport electrons of a given spin
changes sign across the junction.  At finite temperature (bottom panel)
thermal fluctuations of the left and right magnetizations prevent the
layers from maintaining the orientations which maximize the tunnel
resistance, allowing a larger current (yellow arrow) to pass through
the barrier and reducing the TMR ratio.
}
\label{fig:mtj}
\end{figure}

Magnetic tunnel junctions\cite{Julliere1975} (MTJs) have attracted
considerable attention over the past 20 years\cite{Moodera1995,
*Miyazaki1995, *Yuasa2004, *Lee2007, *Chappert2007, Parkin2004} because
their use in read heads\cite{Moser2002, *Mao2006, *Mao2007} has
improved magnetic memory devices, and because they have potential
applications in magnetic switches, magnetic random access
memories,\cite{Daughton1997, *Engel2005, *DeBrosse2004} and in spin
torque\cite{Slonczewski1996, *Berger1996, *Slonczewski1989, *Huai2004,
*Stiles2006, Hayakawa2006} and caloritronic\cite{Walter2011, *Lin2012}
devices.  A MTJ consists of two ferromagnetic layers separated by a
tunnel barrier which is normally not magnetic (see Fig.~\ref{fig:mtj}).
The resistance of a MTJ depends strongly on the relative orientations
of the magnetizations of the two ferromagnetic layers. For most
applications performance is optimized by achieving the largest possible
ambient temperature value of the tunneling magnetoresistance (TMR):
\begin{equation}
\mathrm{TMR}(T) = \frac{G_{P}(T) - G_{A}(T)}{G_{P}(T)} = \frac{R_{A}(T) - R_{P}(T)}{R_{A}(T)}.
\end{equation}
Here, $G_{P}(T)$ ($G_{A}(T)$) and $R_{P}(T)$ ($R_{A}(T)$) are the
conductances and resistances at temperature $T$ for parallel ($P$) and
antiparallel ($A$) magnetization alignments. With this definition a
perfect spin-valve, one in which the tunnel current is completely
shut-off for $A$ alignment, has $\mathrm{TMR}=1$. In order to achieve a
useful device it is important not only that $\mathrm{TMR}(0)$ is close
to one, but also that its $\mathrm{TMR}(T)$ does not decrease
substantially with increasing $T$ up to the intended operation
temperature.  In this article we propose a simple formula for the TMR
temperature dependence, show that it describes the properties of
current high-quality tunnel junctions, and use it to comment on
strategies for achieving better MTJs.

\section{The Stoner Model of $T=0$ TMR}

The magnetic and transport properties of most metallic ferromagnets are
accurately described at $T = 0$ by a mean-field
theory\cite{Gunnarsson1989}, typically the Kohn-Sham equations of
density functional theory, in which majority and minority spin states
experience self-consistently determined spin-dependent potentials. In
the following, we use $\downarrow$ to refer to the majority spin
orientation on the left side of the tunnel barrier. With this
nomenclature, the $\downarrow$ potential is more attractive than the
$\uparrow$ potential on both sides of the barrier for $P$ magnetization
alignment, whereas for $A$ alignment the $\downarrow$ potential is more
attractive on the left side while the $\uparrow$ is more attractive on
the right side.  When spin-orbit coupling is neglected, $\uparrow$ and
$\downarrow$ electrons contribute independently to the conductance in
both cases.  Because the mean-field Hamiltonians are different, the
$\uparrow$ and $\downarrow$ conductances are different in $P$ and $A$
cases, and in the $P$ case also different from each other.
Landauer-B{\"u}ttiker theory\cite{Datta1997} can be reliably applied to
calculate the conductances of accurately characterized tunnel
junctions:
\begin{subequations}
\begin{align}
G_{P} &= G^{MM} + G^{mm}, \\
G_{A} &= G^{Mm}+ G^{mM} = 2 G^{Mm}.
\end{align}
\end{subequations}
Here, the superscript $MM$ refers to majority-spin to majority-spin
tunneling, $mm$ to minority-spin to minority-spin tunneling and $Mm$ to
majority-spin to minority-spin tunneling. In simplified tunneling
Hamiltonian models with spin-independent tunneling amplitudes the
conductance contributed by each spin is proportional to the product of
its density of states on the two sides of the tunnel barrier, so that
$G_{P} \propto \nu_{M}^{2} + \nu_{m}^{2}$, $G_{A} \propto 2 \nu_{M}
\nu_{m}$, and $\mathrm{TMR}(0) > 0$.  Although this model never
strictly applies to real materials, the TMR nevertheless has a strong
tendency to be positive because of the exchange potential jumps across
the junction in the $A$ alignment case. The case of half-metallic
ferromagnets,\cite{Irkhin1994, *Katsnelson2008} in which only
majority-spin states or only minority-spin states are present at the
Fermi level, provides an extreme example because this property implies
that $G^{Mm} = G^{mm} = 0$, and hence perfect spin valve behavior with
$\mathrm{TMR}(0) = 1$.

If the Stoner model captured all relevant physics, the search for
optimal spin valves would reduce to a search for half-metallic
ferromagnets.  Indeed, this search has attracted considerable attention
in the spintronics materials community.\cite{deGroot1983, *Hanssen1986,
*Hanssen1990, *Weht1999, *Lewis1997, *Kobayashi1998, *Gercsi2006,
*Miao2006, *Rybchenko2006, Irkhin1994, *Katsnelson2008}  However,
remarkably high performance spin valves can still be constructed using
materials in which both majority and minority spins are present at the
Fermi level. These junctions generally take advantage of the property
that majority-spin wave functions of elemental transition-metal
ferromagnets decay more slowly than minority-spin wave functions in MgO
and other insulating barrier materials\cite{Butler2001}, causing the
ratio of $G_{P}^{MM}$ to $G_{P}^{mm}$ and $G_{A}^{Mm}$ to grow
exponentially with tunnel barrier thickness within the Stoner model.
The increase in TMR with barrier thickness is eventually limited by
phonon-assisted tunneling and other beyond-Stoner-model effects. It is
nevertheless possible to achieve values of $\mathrm{TMR}(0)$ as large
as $0.879$ in devices with practical values of the tunnel
resistance.\cite{Hayakawa2006}

\section{TMR at finite temperatures: Theory}

As we now explain, uncorrelated thermal fluctuations in magnetization
orientation on opposite sides of the tunnel barrier degrade spin-valve
performance.  This physics is absent in the Stoner model and arises in
formal theoretical analyses from the interaction between quasiparticle
and spin-wave excitations of the ferromagnets.\cite{MacDonald1998}  It
follows that there is a competition in the search for optimal spin
valves between choosing materials that are effectively close to being
half-metallic and choosing materials that have high ferromagnetic
transition temperatures and correspondingly reduced magnetization
orientation fluctuations.

Below we propose a simple approximate expression for the finite
temperature tunnel magnetoresistance $\mathrm{TMR}(T)$ which depends
only on $\mathrm{TMR}(0)$ and on $\zeta(T) = M_{s}(T)/M_{s}(0)$, where
$M_{s}(T)$ is the saturation magnetization.  In the following, we first
present a qualitative discussion which justifies the expression, and
then discuss some of its limitations. The expression is motivated by
the analysis of the one-particle Green's function of an itinerant
electron ferromagnet presented in Ref.~\onlinecite{MacDonald1998}.  As
shown there, at non-zero temperatures both majority-spin and
minority-spin Green's functions have poles at both minority-spin and
majority-spin quasiparticle energies. Provided that the quasiparticle
exchange splitting is larger than spin-wave energies, and that the
temperature is low enough that only long-wavelength spin waves are
thermally excited, the residue of the majority-spin Green's function is
$(1 + \zeta(T))/2$ at majority-spin quasiparticle poles and $(1 -
\zeta(T))/2$ at minority-spin quasiparticles poles.  Similarly, the
residue of the minority-spin Green's function is $(1 - \zeta(T))/2$ at
majority-spin quasiparticle poles and $(1 + \zeta(T))/2$ at
minority-spin quasiparticles poles. The interpretation of these
theoretical results is straight forward. Because of thermal
fluctuations in magnetization orientation an electron with a given
definite spin has a finite probability at finite temperature of being a
majority-spin electron and a finite probability of being a
minority-spin electron.

\begin{figure}[tb]
\includegraphics[width=0.45\textwidth]{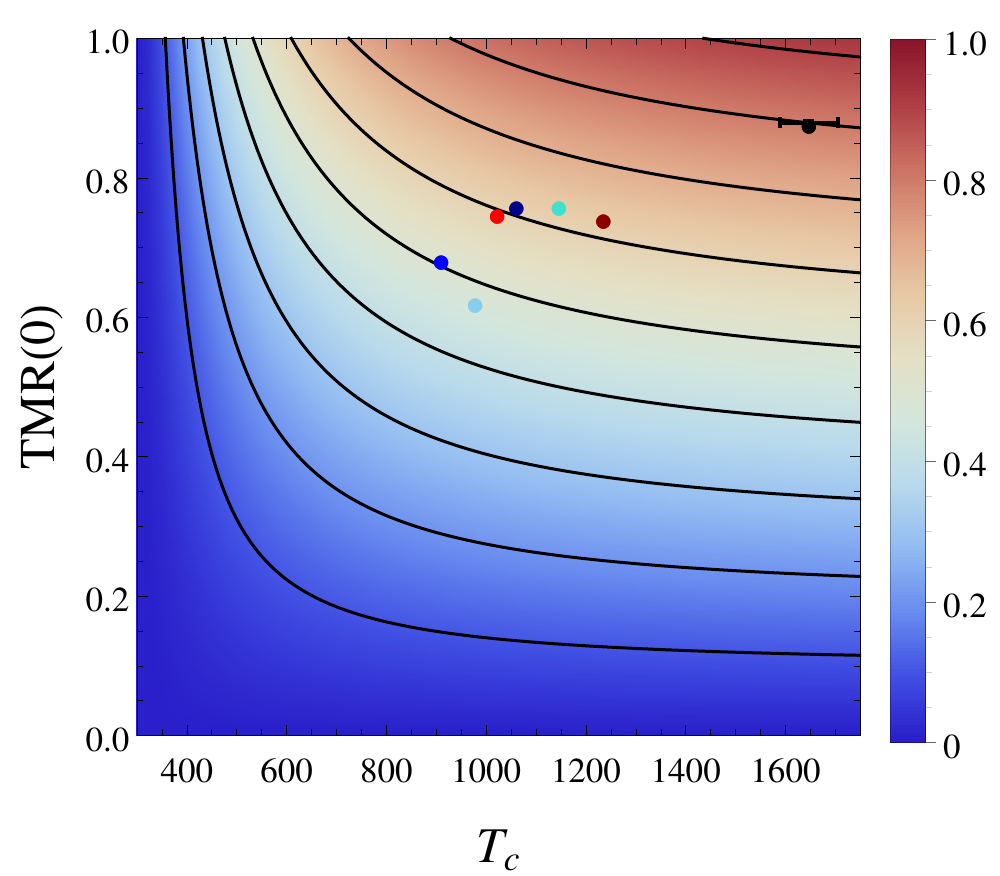}
\caption{%
(Color online) Color-scale plot of
$\mathrm{TMR}(T=\unit{300}{\kelvin})$ as a function of
$\mathrm{TMR}(0)$ and the ferromagnetic critical temperature $T_{c}$
assuming that $\zeta(T) = 1 - (T/T_{c})^{3/2}$.  The solid lines are
contours of constant $\mathrm{TMR}(T)$ separated in value by $0.1$.
The symbols plot the $\mathrm{TMR}_{\mathrm{fit}}(\unit{300}{K})$
values of Tab.~\ref{tab:Experiments} using the same color code as in
Fig.~\ref{fig:DatafitsFe}.  All data points reported as being measured
on the same sample are represented by the same color throughout this
paper.  The error bars represent uncertainties in determining values
from published figures.
}
\label{fig:TMRT}
\end{figure}

When these temperature-dependent quasiparticle weights are included,
many of the deficiencies of the Stoner theory of itinerant electron
magnetism are repaired.  In particular, it is no longer difficult to
reconcile the temperature dependence of the magnetization, which is
controlled by long-wavelength thermally excited magnons whose
occupation numbers are given by the Bose distribution function, with
the temperature dependence of electron quasiparticle occupation numbers
that are given by the Fermi distribution function. In addition, we can
repair the theory of TMR by adding the independent conductivities
contributed by the two average spin orientations and in each case
assigning probabilities for instantaneous spin orientations on each
side of the junction.  For example, the majority-spin conductivity in
the $P$ alignment case is
\begin{widetext}
\begin{equation}
\label{eq:GMMT}
G^{MM}(T) = \lambda(T) \biggl[ G^{MM} \biggl( \frac{1 + \zeta(T)}{2} \biggr)^{2} + G^{mm} \biggl( \frac{1 - \zeta(T)}{2} \biggr)^{2} + 2 G^{Mm} \biggl( \frac{1 + \zeta(T)}{2} \biggr) \biggl( \frac{1 - \zeta(T)}{2} \biggr) \biggr].
\end{equation}
This approximation leads to
\begin{equation}
\label{eq:GPTGAT}
G_{P,A}(T) = \lambda(T) \biggl[ (G^{MM} + G^{mm}) \biggl( \frac{1 \pm \zeta(T)^2}{2} \biggr) + 2 G^{Mm} \biggl( \frac{1 \mp \zeta(T)^{2}}{2} \biggr) \biggr] = \frac{\lambda(T)}{2} \bigl[ G_P \bigl( 1 \pm \zeta(T)^2 \bigr) + G_A \bigl( 1 \mp \zeta(T)^{2} \bigr) \bigr],
\end{equation}
\end{widetext}
where the upper (lower) sign refers to the $P$ ($A$) alignment case.
Hence we find that
\begin{equation}
\label{eq:TMRT}
\mathrm{TMR}(T) =  \mathrm{TMR}(0) \frac{\zeta(T)^{2}}{1 - \mathrm{TMR}(0)(1 - \zeta(T)^{2})/2}.
\end{equation}
The factor of $\lambda(T)$ in Eqs.~\eqref{eq:GMMT} and
\eqref{eq:GPTGAT} accounts for the Fowler-Nordheim\cite{Fowler1928,
*Nordheim1928, Forbes2004, *Forbes2008} thermal smearing effects
responsible for the temperature dependence of tunneling conductances in
non-magnetic systems. At low temperature, $\lambda(T)$ usually
increases quadratically with increasing temperature (see
App.~\ref{app:fitting}). Note that with these definitions $\lambda(0) =
\zeta(0) = 1$. In Fig.~\ref{fig:TMRT} we plot $\mathrm{TMR}(T =
\unit{300}{\kelvin})$ as a function of $\mathrm{TMR}(0)$ and the
ferromagnetic critical temperature $T_{c}$ by assuming that $\zeta(T) =
1 - (T/T_{c})^{3/2}$ to interpolate the dependence of magnetization
between low temperature and the Curie temperature.  Based on this plot
we conclude that in searching for optimal spin-valve behavior a very
high ferromagnetic transition temperature is at least as important as
good low-temperature spin-valve behavior.

Eq.~\eqref{eq:TMRT} becomes exact\cite{MacDonald1998} when (i) there
are no exchange interactions and therefore no correlations between
magnetization orientations across the tunnel barrier, (ii) Boltzmann
weighting factors are large only for magnetization configurations in
which orientations change slowly on a lattice constant length scale,
(iii) thermal fluctuations at the interfaces between nanomagnets and
tunnel barriers are similar to those of bulk magnetic material, and
(iv) the tunneling amplitude across the barrier is spin-independent.
Among these four conditions the first two are normally safely
satisfied.  Because magnetization fluctuations are typically larger at
surfaces of nanomagnets or at interfaces with non-magnetic materials,
the factor $\zeta(T)$ in Eq.~\eqref{eq:TMRT} should likely be chosen to
be smaller than the bulk magnetization thermal suppression factor,
enhancing the importance of a high Curie temperature in achieving good
spin valves.  The fourth requirement for Eq.~\eqref{eq:TMRT} is the
most seriously violated in most TMR systems.  For example,
spin-dependent tunneling figures prominently in yielding the very large
value of $\mathrm{TMR}(0)$ in FeCo/MgO TMR systems.\cite{Butler2001}
Conceptually, $\mathrm{TMR}(T)$ should be performing a
Boltzmann-weighted average of TMR values calculated for all realized
magnetization configurations.  Most configurations have substantial
non-collinearity particularly inside the tunnel barrier.  The main
effect of thermal fluctuations is that non-collinearity increases
$G_{A}$, and thus reduces the TMR.  The property that the typical
degree of local non-collinearity is greater in materials with lower
magnetic transition temperatures is captured by Eq.~\eqref{eq:TMRT}.
The reliability of this equation is discussed further below.

\section{TMR at finite temperatures: Experiment}

\begin{figure*}[tbp]
\includegraphics[width=0.45\textwidth]{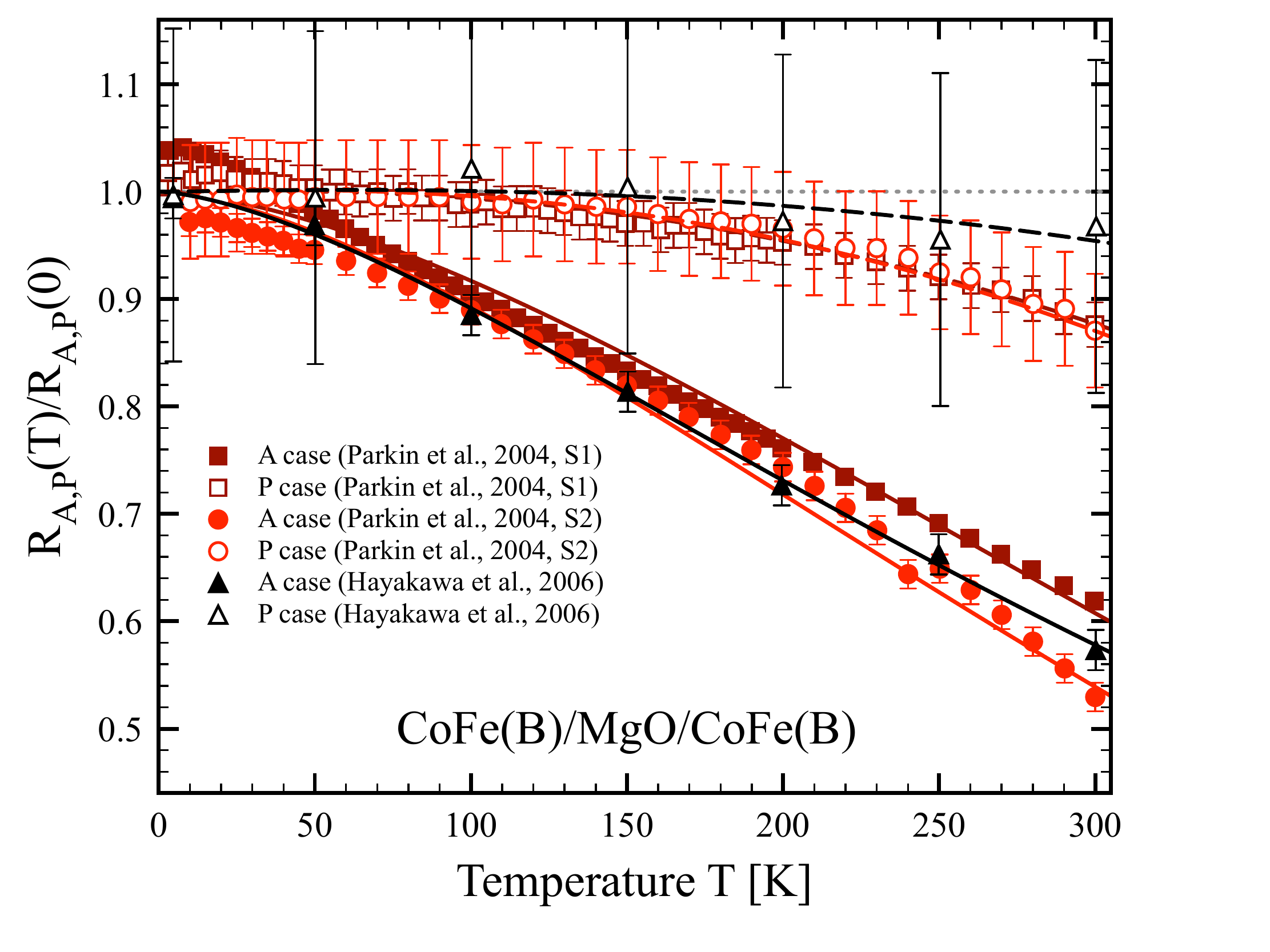}
\includegraphics[width=0.45\textwidth]{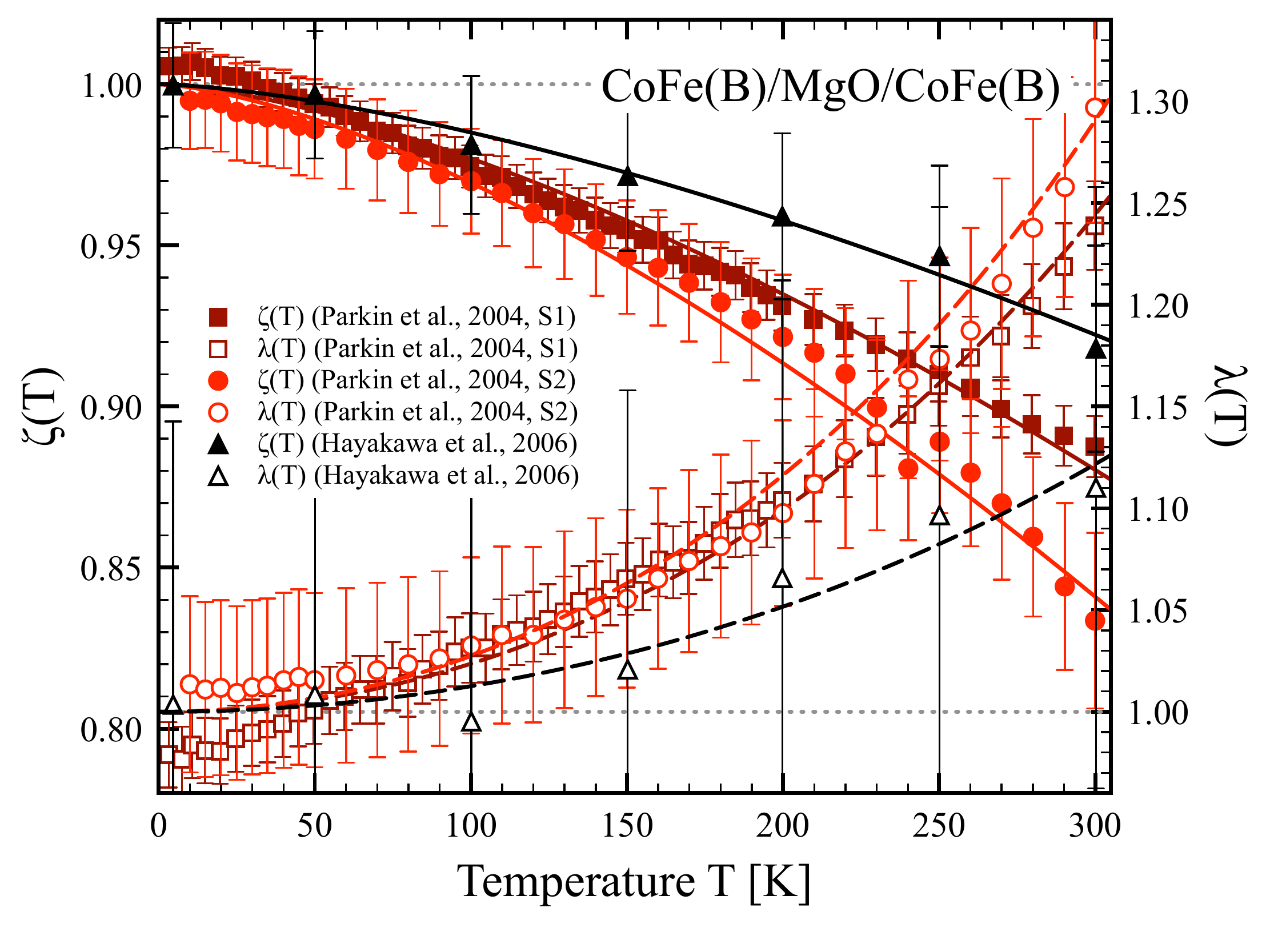}\\
\includegraphics[width=0.45\textwidth]{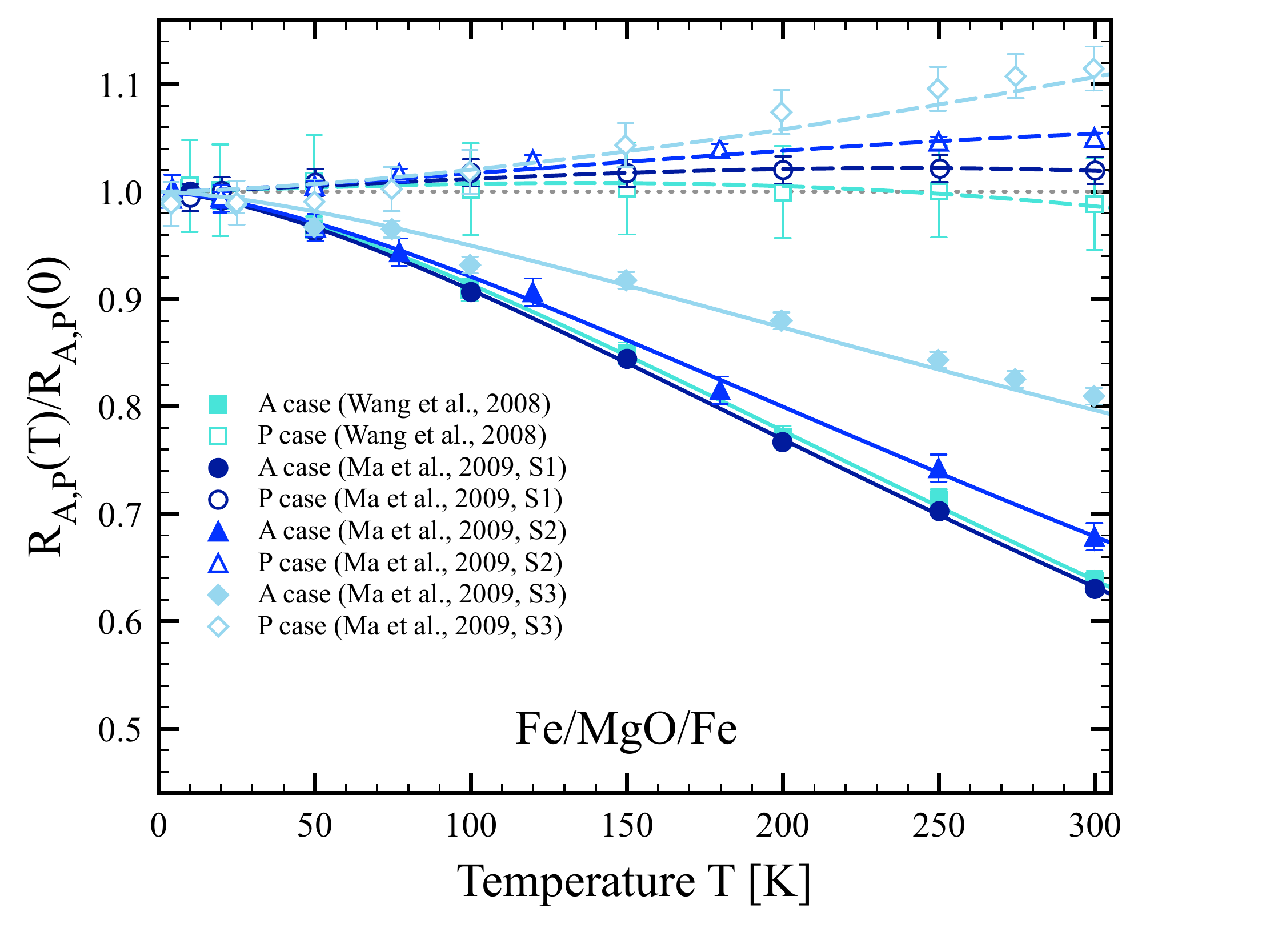}
\includegraphics[width=0.45\textwidth]{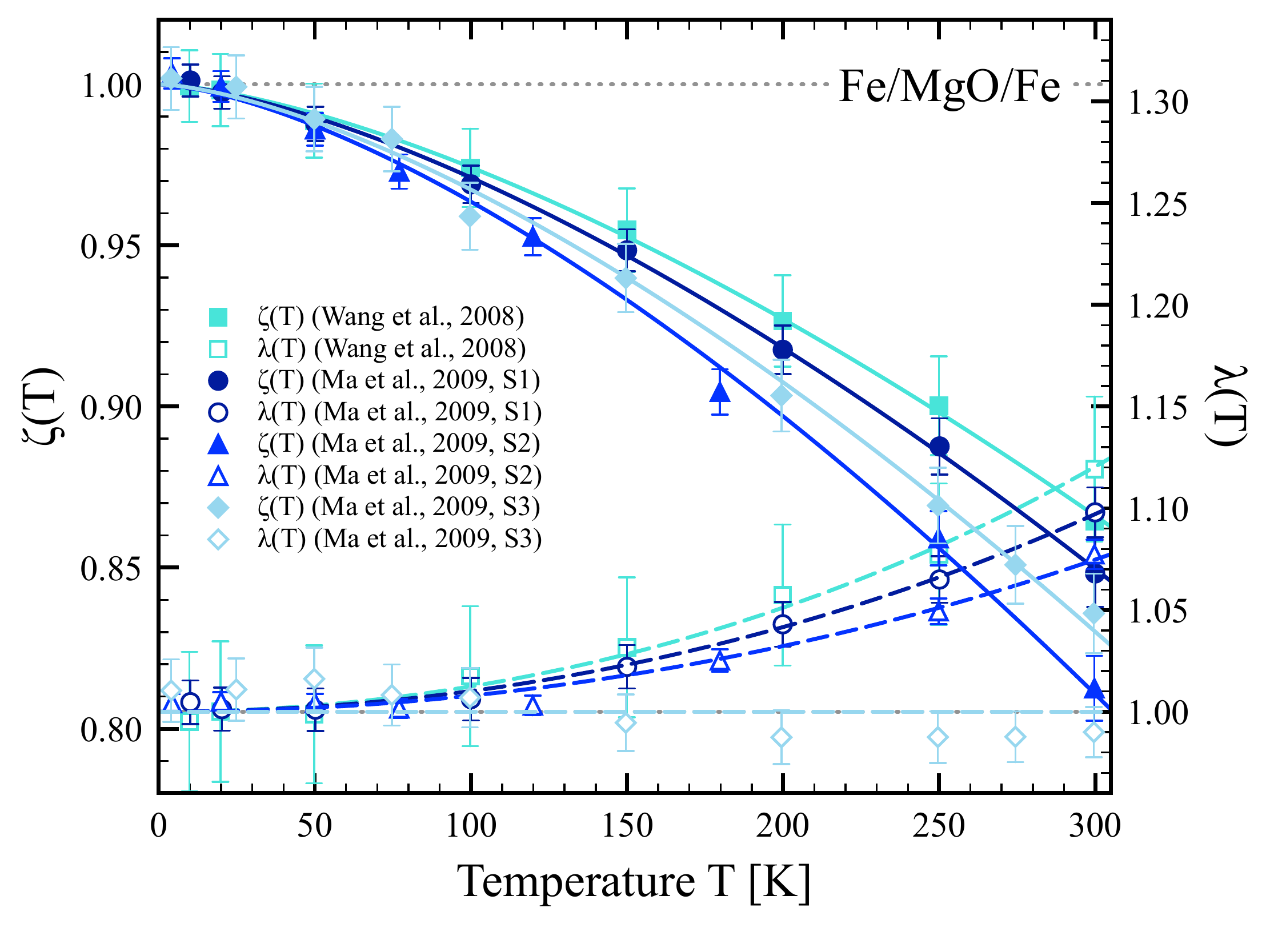}
\caption{%
(Color online) Left column: $R_{P}(T)$ and $R_{A}(T)$ normalized to
their zero-temperature values for the samples studies in Refs.
\citenum{Parkin2004,Hayakawa2006,Wang2008,Ma2009} compared with the
fitting formulas proposed in this paper.  Right column: $\zeta(T)$
(left axis) and $\lambda(T)$ (right axis) fits.  $G_{P}(T) + G_{A}(T)$
increases with temperature in a manner similar to typical behavior for
tunneling between non-magnetic metals. At the same time
$\mathrm{TMR}(T)$ decreases as thermal fluctuations reduce the
dependence of conductance on mean magnetization orientations.  The end
result is that $R_{P}(T)$ ($R_{A}(T)$) is more weakly (strongly)
temperature dependent than in normal metals.  The error bars in these
plots represent uncertainties in determining values from published
figures.
}
\label{fig:DatafitsFe}
\end{figure*}

We have compared recent TMR measurement reports\cite{Parkin2004,
Hayakawa2006, Wang2008, Ma2009} which contains values for both
$G_{P}(T)$ and $G_{A}(T)$ (or equivalently $R_{P}(T)$ and $R_{A}(T)$ or
the corresponding resistance-area products) with Eq.~\eqref{eq:GPTGAT},
forcing agreement by fixing the values of $\lambda(T)$ and $\zeta(T)$:
\begin{subequations}
\label{eq:lambdaandzeta}
\begin{align}
\zeta^{2}(T) &= \frac{G_{P}(T) - G_{A}(T)}{G_{P}(T) + G_{A}(T)} \biggr/ \frac{G_{P}^{0} - G_{A}^{0}}{G_{P}^{0} + G_{A}^{0}}, \\
\lambda(T) &= (G_{P}(T) + G_{A}(T))/(G_{P}^{0} + G_{A}^{0}).
\end{align}
\end{subequations}
In Fig.~\ref{fig:DatafitsFe} we plot experimental data from
Refs.~\citenum{Parkin2004, Hayakawa2006} (top row) and
Refs.~\citenum{Wang2008, Ma2009} (bottom row).  In
Refs.~\citenum{Parkin2004,Hayakawa2006} the authors analyzed
sputter-deposited CoFe(B)/MgO/CoFe(B) MTJs, whereas in
Refs.~\citenum{Wang2008, Ma2009} the authors analyzed epitaxial
MBE-grown Fe/MgO/Fe junctions.  In all these studies, the MgO tunnel
barrier was oriented along (100).  Fig.~\ref{fig:DatafitsFe} presents
fits of this data to Eq.~\eqref{eq:GPTGAT} using $T=0$ conductance and
TMR values, and $T_{c}$ along with a characteristic Fowler-Nordheim
tunneling temperature scale $T_{FN}$ as fitting parameters (for details
see App.~\ref{app:fitting}).  A summary of the MTJ sample geometries,
key data values, and the parameters obtained by fitting the TMR data is
given in  Tab~\ref{tab:Experiments}.  As seen in
Fig.~\ref{fig:DatafitsFe}, the fits generally describe the temperature
dependence of the data well and are most accurate for the MBE-grown
Fe/MgO/Fe samples.

\begin{table*}[tbp]
\centering
\begin{tabular}{l l D{,}{\,\pm\,}{4,2} D{,}{\,\pm\,}{5,5} D{,}{\,\pm\,}{5,5} D{,}{\,\pm\,}{5,5}}
\toprule
  Ref./Sample & MTJ structure & \multicolumn{1}{c}{$T_{c}^{\mathrm{eff}}$ [K]}& \multicolumn{1}{c}{$\mathrm{TMR}_{\mathrm{fit}}(0)$} & \multicolumn{1}{c}{$\mathrm{TMR}_{\mathrm{fit}}(\unit{300}{\kelvin})$} & \multicolumn{1}{c}{$\mathrm{TMR}_{\mathrm{exp}}(\unit{300}{\kelvin})$} \\
\colrule
  Parkin \textit{et al.}, 2004, S1 & 3.0 Co$_{70}$Fe$_{30}$/2.9 MgO/15.0 Co$_{84}$Fe$_{16}$                & 1235,11 & 0.742,0.001 & 0.627,0.001 & 0.635,0.006 \\
  Parkin \textit{et al.}, 2004, S2 & 3.0 Co$_{70}$Fe$_{30}$/3.1 MgO/15.0 Co$_{84}$Fe$_{16}$                & 1022,14 & 0.749,0.002 & 0.595,0.002 & 0.590,0.06 \\
  Hayakawa \textit{et al.}, 2006   & 3.0 Co$_{40}$Fe$_{40}$B$_{20}$/1.5 MgO/3.0 Co$_{40}$Fe$_{40}$B$_{20}$ & 1647,58 & 0.879,0.003 & 0.800,0.003 & 0.81,0.02 \\
\colrule
  Wang \textit{et al.} 2008        & 25.0 Fe/3.0 MgO/10.0 Fe                                               & 1146,10 & 0.761,0.001 & 0.631,0.001 & 0.63,0.02 \\
  Ma \textit{et al.}, 2009, S1     & 25.0 Fe/3.0 MgO/10.0 Fe                                               & 1061,9 & 0.761,0.001 & 0.614,0.001 & 0.630,0.005 \\
  Ma \textit{et al.}, 2009, S2     & 25.0 Fe/2.1 MgO/10.0 Fe                                               &  910,13 & 0.684,0.002 & 0.509,0.002 & 0.512,0.008 \\
  Ma \textit{et al.}, 2009, S3     & 25.0 Fe/1.5 MgO/10.0 Fe                                               &  978,14 & 0.622,0.002 & 0.474,0.002 & 0.477,0.009 \\
\botrule
\end{tabular}
\caption{%
Comparison of experimental room temperature TMR data from Refs.~\citenum{Parkin2004, Hayakawa2006, Wang2008, Ma2009} to
our theory, Eq.~\eqref{eq:TMRT}.  The stack layer thicknesses are in nm units. The errors stem from uncertainties in determining values from published figures.
}
\label{tab:Experiments}
\end{table*}

\begin{figure*}[tbp]
\includegraphics[width=0.45\textwidth]{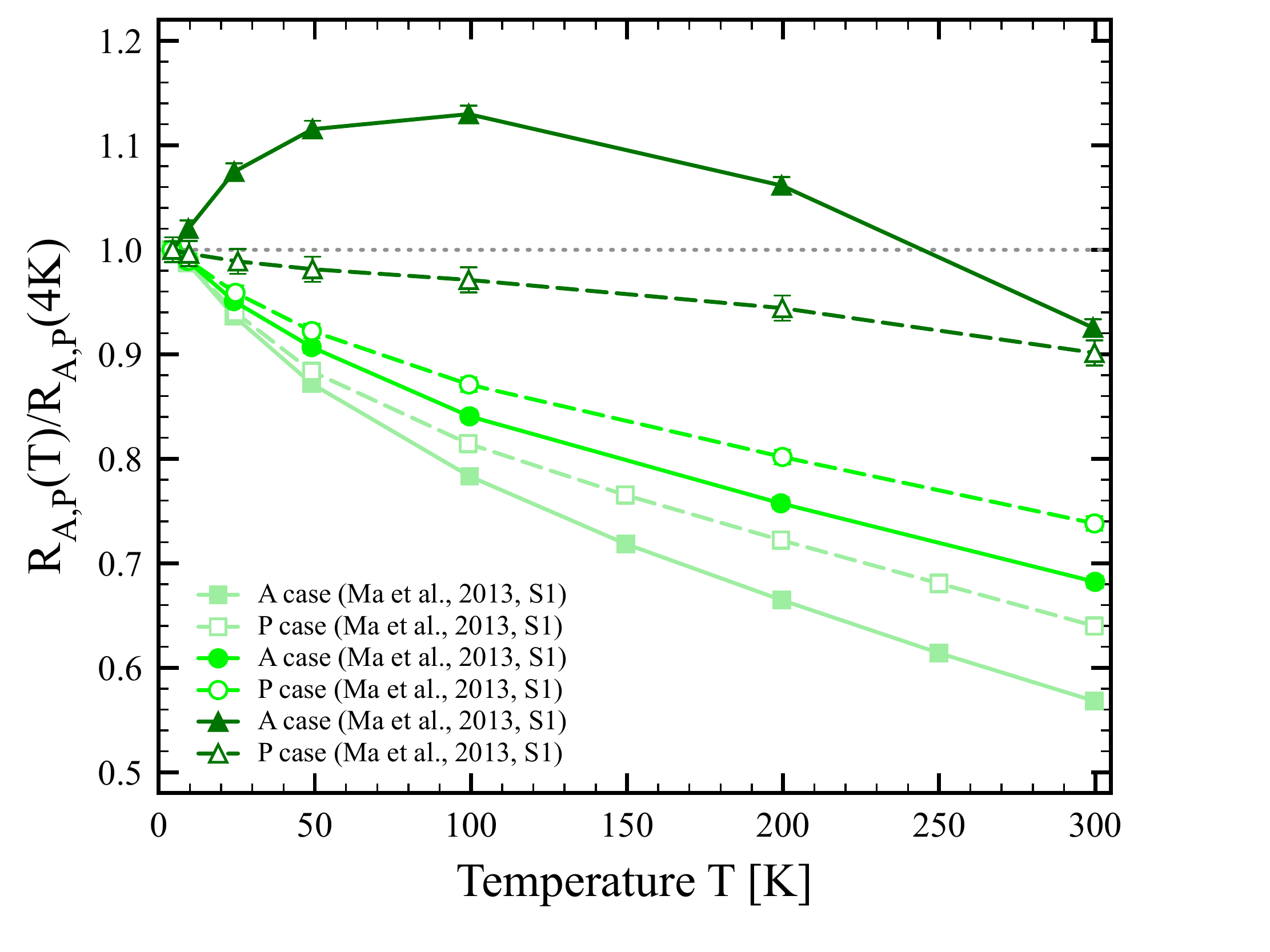}
\includegraphics[width=0.45\textwidth]{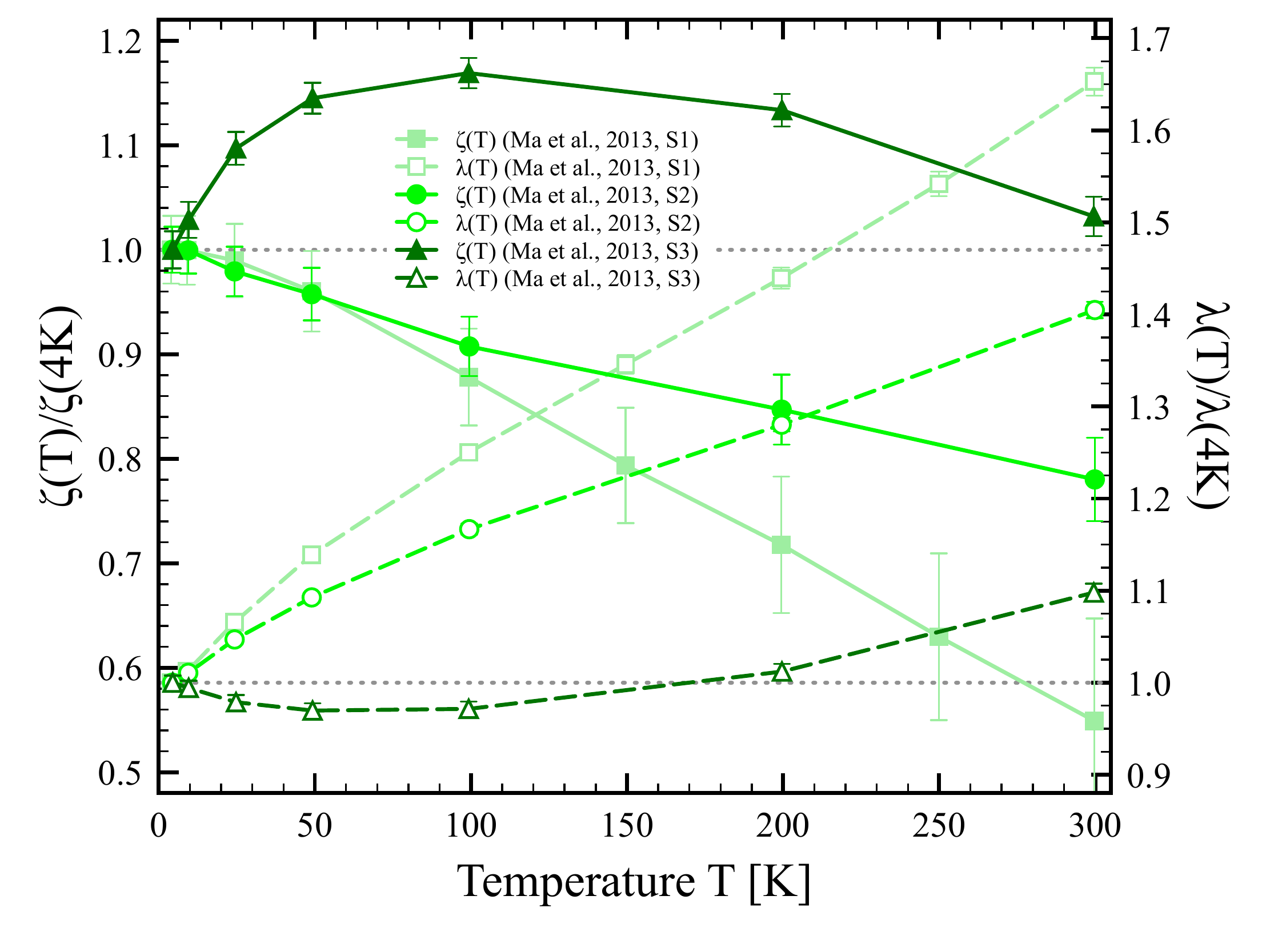}
\caption{%
(Color online) Left panel: $P$ and $A$ resistances \textit{vs.}\
temperature for the data of Ref.~\citenum{Ma2013}. Right panel:
$\zeta(T)$ (left axis) and $\lambda(T)$ (right axis) calculated from
the data using Eq.~\eqref{eq:lambdaandzeta} and normalized to their
values at $T = \unit{4}{\kelvin}$. For sample S1:
$\mathrm{t}_\mathrm{CoFeB}=\unit{0}{\nano\meter}$, S2:
$\mathrm{t}_\mathrm{CoFeB}=\unit{1.0}{\nano\meter}$, S3:
$\mathrm{t}_\mathrm{CoFeB}=\unit{1.5}{\nano\meter}$.  The error bars
represent uncertainties in determining values from published figures.
}
\label{fig:Datafits2}
\end{figure*}

It is important to observe that there are data in the literature which
are not well described by our theory.  One example are the observations
reported in Ref.~\citenum{Ma2013}, summarized in
Fig.~\ref{fig:Datafits2}, which studied 30
Mn$_x$Ga$_{100-x}$/$\mathrm{t}_{\mathrm{CoFeB}}$CoFeB/0.4 Mg/2.2
MgO/0.2 Mg/1.2 CoFeB MTJ stacks.  The thickness
$\mathrm{t}_{\mathrm{CoFeB}}$ of the CoFeB layer adjacent to the
(Mn,Ga)  was varied between $0$ and $\unit{1.5}{\nano\meter}$.  Our
theory completely fails to describe the two samples with
$\mathrm{t}_{\mathrm{CoFeB}}=\unit{0}{\nano\meter}$ and
$\mathrm{t}_{\mathrm{CoFeB}}=\unit{1.0}{\nano\meter}$ for which the
influence of the Mn$_x$Ga$_{100-x}$ layer is largest.  The authors
state that Mn atoms in these samples diffuse into the MgO layer.  It
seems likely that the Mn atoms induce local moments in the MgO that are
detrimental for TMR and lie outside the physics that our approximate
formula aims to capture.  For the sample in this series with the
thickest CoFeB, $\mathrm{t}_{\mathrm{CoFeB}}=\unit{1.5}{\nano\meter}$,
curves in Fig.~\ref{fig:Datafits2} are more similar to those in
Fig.~\ref{fig:DatafitsFe} above about $\unit{100}{\kelvin}$.

\section{Summary and Conclusions}
\label{sec:Conclusions}

We have proposed a simple formula, Eq.~\eqref{eq:TMRT}, for the
temperature-dependent TMR of a MTJ.  Eq.~\eqref{eq:TMRT} captures the
property that the effectiveness of a tunnel-junction spin-valve depends
both on the way in which spin-dependent exchange potentials influence
the transport properties of electronic quasiparticles driven through a
tunnel junction by a bias voltage, and on the degree to which those
exchange potentials thermally fluctuate around mean values which can be
controlled experimentally.  The physics which controls the value of the
TMR at $T=0$ depends mainly on electronic structure considerations in
magnetic materials which are often accurately described by spin-density
functional theory.  For specific material combinations
$\mathrm{TMR}(0)$ can be calculated accurately by separately evaluating
the tunnel conductances of $P$ and $A$ magnetization configurations
using the Landauer-B{\"u}ttiker formula\cite{Datta1997} and the Green's
function methods of nanoelectronics. $\mathrm{TMR}(0)$ reaches its
maximum value when either only majority or only minority spins are
present at the Fermi level, \textit{i.e.}\ in the case of half-metallic
ferromagnetism.  However, for sufficiently thick barriers quite large
values of $\mathrm{TMR}(0)$ can still be achieved when both minority
and majority spins have substantial presence at the Fermi energy,
provided that the wave functions of one or the other decay more slowly
in the tunnel barrier. At finite temperatures electronic structure
theory considerations do not capture the most important source of TMR
temperature dependence, uncorrelated thermal fluctuations in
magnetization orientation on opposite sides of the tunnel barrier. This
effect is captured in a very simple way in Eq.~\eqref{eq:TMRT}, where
$\mathrm{TMR}(T)$ is related to the degree to which thermal
fluctuations reduce the magnetization close to the tunnel barrier,
$\zeta(T) = M_{s}(T)/M_{s}(0)$.  In order to interpret literature
experimental data using Eq.~\eqref{eq:TMRT}, we have assumed that in
any viable spin-valve system we can use the low-temperature expression
$\zeta(T) = 1 - (T/T_{c})^{3/2}$ for $\zeta(T)$.  The $T_{c}$ values we
obtain by fitting literature TMR data to Eq.~\eqref{eq:TMRT}, listed in
Tab.~\ref{tab:Experiments}, should be understood not literally as
ferromagnetic critical temperature values, but as a number which
characterizes the relative strength of thermal fluctuation effects in
particular MTJ systems.  The fact that the TMR fitting temperatures are
nevertheless in each case comparable to the true ferromagnetic critical
temperatures associated with the employed materials ($T_{c} \sim
\unit{1040}{\kelvin}$ in bulk Fe, $\sim \unit{1400}{\kelvin}$ in bulk
Co), provides strong evidence that we have correctly identified the
origin of temperature dependence. In the light of this agreement, we
are confident that Eq.~\eqref{eq:TMRT} can be used to estimate the room
temperature TMR of high-quality spin-valve systems given only their
$\mathrm{TMR}(0)$ and the magnetic critical temperature of its magnetic
constituents (see Fig.~\ref{fig:TMRT}).  We can therefore conclude that
the room temperature TMR values of spin-valves fabricated with
half-metallic magnetic materials can exceed those of existing
spin-valves fabricated with Fe, Co, CoFeB, and similar magnetic
materials combined with MgO tunnel barriers only if the half-magnetic
material has a critical temperature exceeding at least $\sim
\unit{900}{\kelvin}$.  The highest reported room temperature TMR value
with standard materials is $\sim 0.81$ (Ref.~\citenum{Hayakawa2006}).
According to Eq.~\eqref{eq:TMRT}, this value is achieved for
$\mathrm{TMR}(0)=1$ if $\zeta(\unit{300}{\kelvin}) = 0.825$ which corresponds to
$T_{c}^{\mathrm{eff}} = \unit{960}{\kelvin}$. These
considerations rule clearly out most known half-metallic magnetic
materials, with ferrimagnetic Fe$_{3}$O$_{4}$ ($T_{c} =
\unit{858}{\kelvin}$) standing out as a notable exception.

\section*{Acknowledgments}

This work was supported by the DOE Division of Materials Sciences and
Engineering under grant DE-FG03-02ER45958 and by the Welch foundation
under grant TBF1473. KE and MS acknowledge financial support from DAAD.

\appendix

\section{Fitting procedure}
\label{app:fitting}

We extracted resistance (or resistance-area product) data $R_{A,P}(T)$
from Refs.~\citenum{Parkin2004, Hayakawa2006, Wang2008, Ma2009}.
Starting from Eq.~\eqref{eq:GPTGAT} in the main text, we can express
$\zeta(T)$ and $\lambda(T)$ in terms of the measured resistances using
\begin{subequations}
\begin{align}
\zeta^{2}(T) &= \frac{R_A(T) - R_{P}(T)}{R_A(T) + R_P(T)} \biggr/ \frac{R_A(0) - R_{P}(0)}{R_A(0) + R_P(0)}, \\
\lambda(T) &= \frac{R_A(T) + R_{P}(T)}{R_A(T) R_P(T)} \biggr/ \frac{R_A(0) + R_{P}(0)}{R_A(0) R_P(0)}
\end{align}
\end{subequations}
with $\lambda(0)=\zeta(0)=1$.  As discussed in the main text, we expect
the temperature dependence of the saturation magnetization behaviour to
fall off in temperature as $\zeta(T)=1-(T/T_{c}^{\mathrm{eff}})^{3/2}$
for $T < T_{c}^{\mathrm{eff}}$ and fit results at all temperatures
using a single parameter $T_{c}^{\mathrm{eff}}$ which reflects the
strength of thermal magnetization fluctuations in the MTJ device.
$\lambda(T)$ accounts for the Fowler-Nordheim thermal smearing
effects\cite{Fowler1928, *Nordheim1928, Forbes2004, *Forbes2008} and is
commonly modeled using $\lambda(T)=  [ ( T/T_{\mathrm{FN}})/ \sin(
T/T_{\mathrm{FN}})]\approx 1+ (T/T_{\mathrm{FN}})^{2}/6$, where we have
introduced a Fowler-Nordheim temperature $T_{\mathrm{FN}}$ defined by
$T_{\mathrm{FN}}= d_{F}/(\pi k_B)$.  $d_{F}$ is the barrier decay
length expressed in energy units and can be directly related to the
barrier height.\cite{Forbes2004, Forbes2008}  We have used one
parameter $T_{FN}$ to fit the measured resistances at all temperatures.

Because zero-temperature data are not always available we have found it
convenient to perform instead a closely related four-parameter fit of
the temperature-dependent $P$ and $A$ resistances by defining
\begin{subequations}
\begin{align}
\tilde{\zeta}(T) &= \biggl( \frac{R_{A}(T) - R_{P}(T)}{R_{A}(T) + R_P(T)} \biggr)^{1/2}, \\
\tilde{\lambda}(T)&= \frac{R_{A}(T) + R_{P}(T)}{R_{A}(T) R_{P}(T)},
\end{align}
\end{subequations}
and fitting measured values to the following functions:
\begin{subequations}
\begin{align}
\tilde{\zeta}(T) &= \tilde{\zeta}_{0} \bigl[ 1 - (T/T_{c}^{\mathrm{eff}})^{3/2} \bigr], \\
\tilde{\lambda}(T)&= \tilde{\lambda}_{0} \bigl[ (T/T_{\mathrm{FN}}) / \sin(T/T_{\mathrm{FN}}) \bigr].
\end{align}
\end{subequations}
The values of the four parameters $ T_{c}^{\mathrm{eff}},
T_{\mathrm{FN}},\tilde{\zeta}_{0}, \tilde{\lambda}_{0}$ which provide the
best fits to the data shown in Fig.~\ref{fig:DatafitsFe} are listed in
Tab.~\ref{tab:Experimentsdetails}.  From these fit parameters we
obtain $\mathrm{TMR}(0)$ using:
\begin{subequations}
\begin{align}
R_{P,A}(0) &= 2/[\tilde{\lambda}_{0} (1 \pm \tilde{\zeta}_{0}^{2})], \\
\mathrm{TMR}(0) &= 2 \tilde{\zeta}_{0}^{2}/(1 + \tilde{\zeta}_{0}^{2}).
\end{align}
\end{subequations}

\begin{table}[tb]
\centering
\begin{tabular}{l c c c c}
\toprule
  Ref./Sample & $T_{c}^{\mathrm{eff}}$ [K] & $T_{FN}$ [K] & $\tilde{\zeta}_{0}$ & $\tilde{\lambda}_{0}$ [$\Omega^{-1}$] \\
\colrule
  Parkin \textit{et al.}, 2004, S1  & 1235 & 268 & 0.768 & $6.55 \cdot 10^{-3}$ \\
  Parkin \textit{et al.}, 2004, S2  & 1022 & 249 & 0.774 & $1.655 \cdot 10^{-2}$ \\
  Hayakawa \textit{et al.}, 2006    & 1647 & 366 & 0.885 & $2.90$ \\
\colrule
  Wang \textit{et al.} 2008         & 1146 & 367 & 0.784 & $3.781 \cdot 10^{-5}$ \\
  Ma \textit{et al.}, 2009, S1      & 1061 & 406 & 0.784 & $7.83 \cdot 10^{-7}$ \\
  Ma \textit{et al.}, 2009, S2      &  910 & 459 & 0.721 & $7.58 \cdot 10^{-5}$ \\
  Ma \textit{et al.}, 2009, S3      &  978 & $\infty$ & 0.672 & $1.129 \cdot 10^{-3}$ \\
\botrule
\end{tabular}
\caption{%
Fit parameters for data from Refs.~\citenum{Parkin2004, Hayakawa2006,
Wang2008, Ma2009}.
}
\label{tab:Experimentsdetails}
\end{table}

\bibliography{mtj}

\end{document}